\documentclass[12pt]{emulateapj}

\shorttitle{Kinematics of Simulated Galaxies}
\shortauthors{Kassin et al.}

\usepackage{soul}
\usepackage{amsmath}

\begin{document}

\title{Kinematic Evolution of Simulated Star-Forming Galaxies}

\author{Susan A. Kassin,\altaffilmark{1}
Alyson Brooks,\altaffilmark{2}
Fabio Governato,\altaffilmark{3}
Benjamin J. Weiner,\altaffilmark{4} and
Jonathan P. Gardner\altaffilmark{5} 
}
\altaffiltext{1}{Space Telescope Science Institute, 3700 San Martin Drive, Baltimore, MD 21218, kassin@stsci.edu}
\altaffiltext{2}{Department of Physics and Astronomy, Rutgers University, 136 Frelinghuysen Road, Piscataway, NJ 08854}
\altaffiltext{3}{Astronomy Department, University of Washington, Box 351580, Seattle, WA 98195-1580}
\altaffiltext{4}{Steward Observatory, 933 N. Cherry St., University of Arizona, Tucson, AZ 85721}
\altaffiltext{5}{Astrophysics Science Division, Goddard Space Flight Center, Code 665, Greenbelt, MD 20771}

\begin{abstract}
Recent observations have shown that star-forming galaxies like our own Milky Way evolve kinematically into 
ordered thin disks over the last $\sim 8$ billion years since $z=1.2$, undergoing a process of ``disk settling."  
For the first time, we study the kinematic evolution of a suite of 
four state of the art ``zoom in" hydrodynamic simulations of galaxy formation and evolution in a fully cosmological 
context and compare with these observations.  Until now, robust measurements of the internal kinematics of 
simulated galaxies were lacking as the simulations suffered from low resolution, overproduction of stars, 
and overly massive bulges.  The current generation of simulations has made great progress in overcoming these difficulties
and is ready for a kinematic analysis.
We show that simulated galaxies follow the same kinematic trends as real galaxies: 
they progressively decrease in disordered motions ($\sigma_g$) and increase in ordered rotation ($V_{rot}$) with 
time.  The slopes of the relations between both $\sigma_g$ and $V_{rot}$ with redshift are consistent between the
simulations and the observations.  In addition, the morphologies of the simulated galaxies become
less disturbed with time, also consistent with observations, and they both have similarly large scatter.
This match between the simulated and observed trends is a significant success for the 
current generation of simulations, and a first step in determining the physical processes behind disk settling.
\end{abstract}


\section{Introduction}

Over the last $\sim 8$ billion years since a redshift of one, the population of star-forming galaxies of $\sim$Milky Way mass has
settled kinematically into flat, rotationally-supported, disk galaxies \citep{kas7, kas12}.  In the past, these galaxies 
had larger integrated gas velocity dispersions ($\sigma_g$) and 
less ordered rotation ($V_{rot}$) than they do today \citep{flor, kas7, verg, kas12}.  
This $\sigma_g$ is likely due to disordered motions in these galaxies \citep{covi, kas12}.
Over $0.1<z<1.2$, the median $\sigma_g$ of star-forming galaxies of $\sim$Milky Way mass has progressively decreased while $V_{rot}$ has increased
\citep{kas12}.  Observations at even higher redshifts ($z\sim 1.5 -3$), albeit where galaxy
samples are much less representative, also find star-forming galaxies with large amounts of disordered motions,
as measured through $\sigma_g$ \citep[e.g.,][]{fors9, law, wright9, lem15, lem3, gner}.

These findings are an important benchmark for any theory or simulation of disk galaxy formation.  Constraints on 
the internal kinematics of galaxies are significantly more stringent than constraints on e.g., luminosity, 
stellar mass, or star formation rate.  Internal kinematics directly dictate how the gas in galaxies is arranged,
which is heavily influenced by the physical processes involved in galaxy evolution.
Furthermore, if theory is able to successfully reproduce observations of kinematic evolution, then it becomes a useful tool to help interpret
the observations.  In particular, we would like to know the physical processes behind disk settling.

Until now, robust measurements of the internal kinematics of simulated galaxies were lacking
as the simulations suffered from low resolution, overproduction of stars, and overly massive bulges.
These problems were ameliorated
by increased resolution \citep[e.g.,][]{maye}, better treatment of feedback \citep{stin, hopk12, wise},
and modeling of the ultraviolet background \citep{quin}.  
Among other things, the current generation of simulations are better able to model star-formation, inflows, and outflows than
previous generations. This makes them better able to match the rotation velocities, sizes, stellar masses, and baryon fractions of local disk
galaxies \citep[e.g.,][]{brooks11, brook12, muns, chri2, vogel}, as well as some properties
of higher redshift galaxies \citep[e.g.,][]{hirs,zemp,hopk}.
These simulations are the first generation which are able to make quantitative predictions for the 
internal kinematics of galaxies.  

A few studies using these current generation simulations have measured kinematics, and most of them have
only measured rotation velocities for local disk galaxies \citep[e.g.,][]{gued, stin13, mcca, brook12}.  
Fewer simulations have measured rotation velocities at higher redshifts, and fewer still have measured $\sigma_g$
at any redshift: e.g., \citet{rok} and \citet{ager} for $z=0$, \citet{croft} for $z=0$ and $z=1$,  \citet{ceve} and \citet{aa} for $z=2$,
and \citet{robe} which studied gas-rich mergers which should be common at $z=2$.
\citet{bird} studied the evolution of $V_{rot}/\sigma_g$ over the lifetime of a single star-forming galaxy.
 No studies have yet compared simulated galaxies to the progressive disk settling found at $0.1<z<1.2$ by \citet{kas12}.
In this paper, for the first time, we determine whether a suite of fully cosmological simulations of star-forming galaxies undergoes the kinematic 
disk settling found in observations of real galaxies.

\begin{figure*}
\includegraphics[scale=1.36]{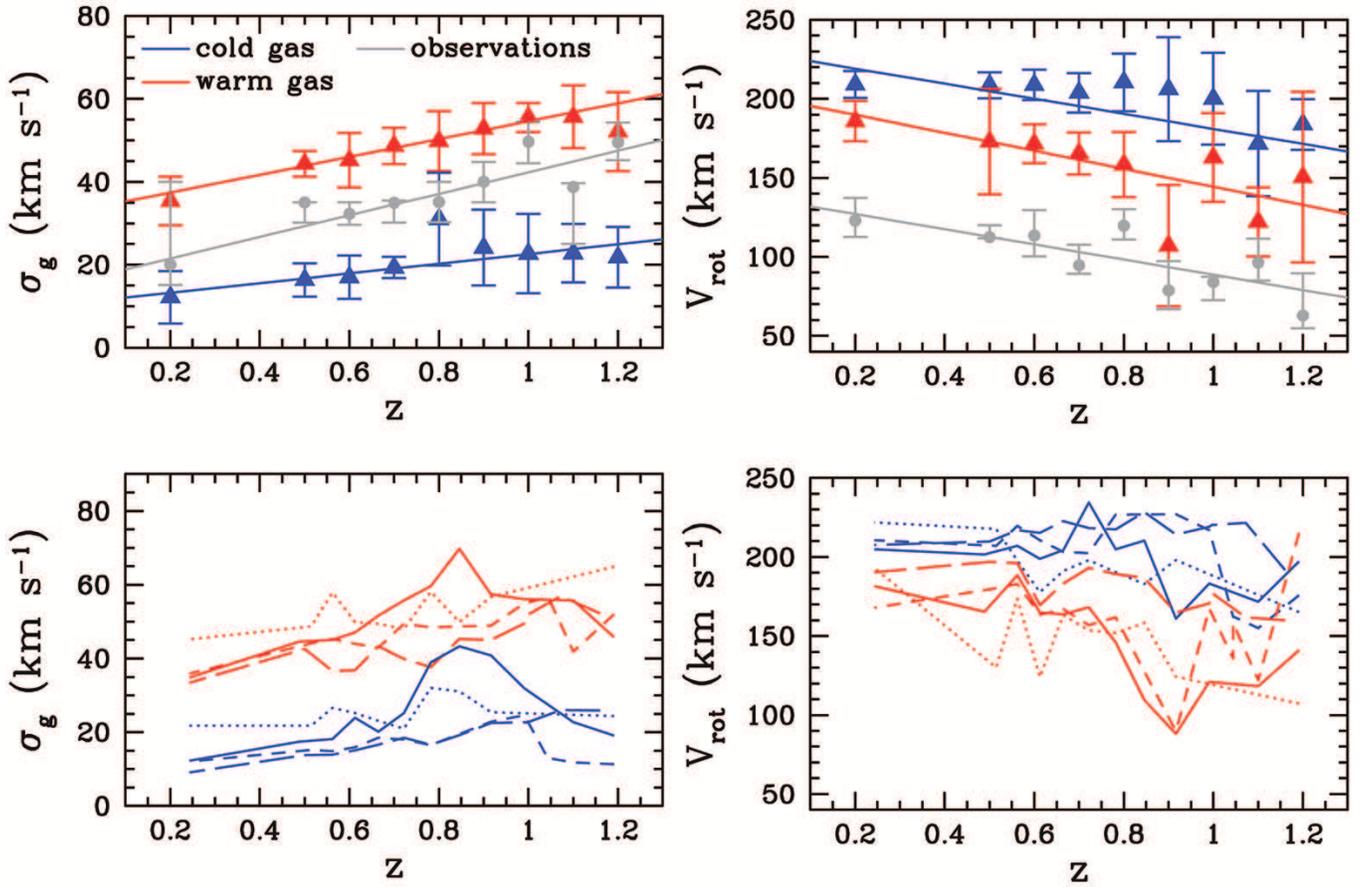}
\caption{{\sf Top:} The redshift evolution of the median integrated velocity dispersion $\sigma_g$ and rotation velocity $V_{rot}$ 
of warm and cold gas in the simulated galaxies (red and blue points, respectively) is compared 
with that for an observational mass-limited sample of 270 galaxies from \citealt{kas12} (grey points).   
The simulations follow the same disk settling trends as the observations: they increase in $\sigma_g$
and decrease in $V_{rot}$ with redshift (i.e., decrease in $\sigma_g$ and increase in $V_{rot}$ with time).
The differences in normalizations between the simulations and observations can be attributed
to differences in galaxy stellar masses for $V_{rot}$ and temperatures/regions probed for $\sigma_g$ (see text).
Solid lines show linear fits given in the text.   Error bars are calculated differently for the simulations and
the observations.  Error bars on the simulations show the rms scatter.
As in \citealt{kas12}, error bars on the observations are calculated by bootstrap re-sampling the data in each 
redshift bin, since their distributions are non-Gaussian.
{\sf Bottom:}  The evolution of the warm and cold gas (red and blue lines, respectively) in the four
individual simulated galaxies is shown: h285 (solid), h239 (dotted), h258 (short dashed), and h277 (long dashed).  
Although the simulated galaxies show the same trends as the observations in the median, individually they
show significant variation with time, consistent with the large scatter of the observations.
\label{settling}}
\end{figure*}

\begin{figure*}
\includegraphics[scale=2.42]{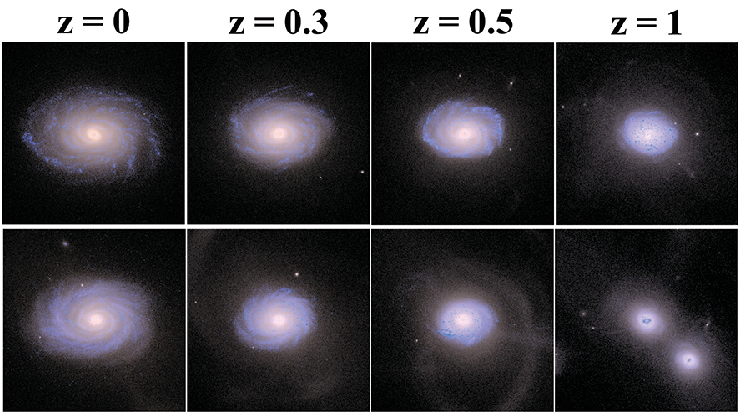}
\caption{Multi-color images of two of our four simulated galaxies are shown at discrete redshifts spanning the redshift range of Figure~\ref{settling}.
The morphologies become progressively less disturbed with time and the disk grows (from right to left), 
reflective of the kinematic evolution which shows disk settling with time.
The images are for galaxies h277 (top panels) and h239 (bottom panels) and are combinations of $g,r$, and $i$-band 
images created with the {\tt Sunrise} radiative transfer code.  They are 50 kpc (physical) on a side 
and the central galaxy in each image is viewed at an inclination of 45\degr.
\label{images}}
\end{figure*}
 
\section{Hydrodynamic Simulations}

The kinematics of star-forming galaxies in the real universe are measured from nebular emission lines which trace 
gas heated in star-forming regions.  In order to compare hydrodynamic simulations of galaxies most directly 
with observations, the simulations need to make predictions for the gas in galaxies.
Treatment of the interstellar medium is crucial for this.  Therefore,
we look to simulations of four star-forming galaxies run with the N-Body + smooth particle hydrodynamic code
GASOLINE \citep{wads, stin} which are described in detail in \citet{chri1,chri2} and \citet{muns}.  
They are able to match the sizes, stellar masses, bulge masses, and baryon fractions of local star-forming galaxies \citep{gov7, brook12, muns, chri2},
and the size evolution of star-forming galaxies since $z=1$ \citep{brooks11}.
The simulated galaxies have $z=0$ stellar masses which range $4.2-4.5\times 10^{10}$ M$_{\odot}$ for a \citet{krou} initial mass function (IMF).
They are referred to as galaxies h239, h277, h258, and h285 in the references above. 

These simulations resolve high density peaks comparable in size
to giant molecular clouds.  In addition, they follow the 
non-equilibrium formation and destruction of H$_2$, using a gas-phase and a 
dust- (and hence metallicity-) dependent scheme that traces the Lyman-Werner radiation 
field and allows for self-shielding by H$_2$ gas \citep{gned, chri2}.  
Therefore, the simulations are able to tie star formation to the presence of molecular gas, as is 
indicated by observations \citep[e.g.,][]{lero, bigi8, blan, bigi10, schr}. 
The efficiency of star formation is tied to the H$_2$ fraction, as described in \citet{chri1}.
With the inclusion of H$_2$-based star formation, stars predominantly form at high 
densities and low temperatures \citep[T $< 1000$ K; see also][]{krum, 
kuhl}.  Finally, to take into account the effects of the reionization of the universe, an ultraviolet background 
is implemented at $z=9$, following a modified version of the formulation by \citet{haa}. 

Our simulated galaxies were chosen from lower resolution simulations to be re-simulated at high resolution
based on their $z=0$ virial masses and because they span a representative range of halo spin values and accretion histories.
The re-simulations follow a 50 Mpc co-moving box around the galaxies at lower resolution than the galaxy itself.
In this manner, cosmological effects are fully implemented.   The re-simulations are run from $z=150$ to $z=0$.
 
The spline force softening in the high resolution region is 174 parsecs, and is
kept fixed in physical parsecs since $z=10$.  In the high resolution region, dark matter particles have masses of $1.3\times 10^5 M_{\odot}$,
gas particles start at $2.7\times 10^4 M_{\odot}$, and star particles are born with 30\% of the mass of their parent gas
particle, which corresponds to a maximum initial mass of $8100\,M_{\odot}$.  Each simulated galaxy has $\sim5$ million
dark matter particles within its virial radius at $z=0$ and more than 14 million total particles (dark matter, gas, and stars).
Our simulations have the same mass resolution as the Eris simulation \citep{gued}, but include metal line 
cooling \citep{shen} and the physics of molecular hydrogen \citep{chri1}.   
The simulations are run for a WMAP 3 Year Cosmology: $\Omega_m = 0.24, \Lambda = 0.76, \rm H_o = 73$ km s$^{-1},$ 
$\sigma_8 = 0.77$ \citep{sper}.

\subsection{Warm \& Cold Gas in the Simulations}

The nebular emission lines in real galaxies, from which internal kinematics are measured, largely come from \ion{H}{2} regions, with a small 
contribution from diffuse ionized gas. Gas in \ion{H}{2} regions has been ionized by young stars after the stars and gas emerge from their 
dusty birth environments inside dense regions of molecular clouds.  We investigate two types of star forming gas in the simulations which 
should bracket the kinematic behavior of \ion{H}{2} regions: (1) cold gas in the galaxy disks which we refer to as ``cold gas", and (2) dense 
gas in the galaxy disks which has been heated by supernovae feedback and which we refer to as ``warm gas." Cold gas is defined as having 
temperatures $T < 1000$K. It has typical densities of $> 1$ amu cm$^{-3}$ and traces both neutral gas in the disk and molecular clouds at $\rho > 100$ amu
cm$^{-3}$ \citep{chri2}. Warm gas is defined as having $\rho >  0.001$ amu cm$^{-3}$); it has temperatures in excess of 20,000K due to heating by feedback. 
Star-forming regions in observed galaxies likely contain ionized gas from both the cold and warm components because they are being 
observed after the young stars are visible and some feedback has had a chance to occur. Therefore, we expect the behavior of the cold and 
warm gas in the simulations to bracket that of HII regions.  However, we note that the current generation of simulations
still has a problem with keeping cold gas in galaxies without forming stars.  Since the cold gas does not have a
long lifetime, the feedback-affected gas dominates the gas mass at all times.

\section{Kinematic Evolution of Simulated Galaxies}

For each of the 4 simulated galaxies, we measure $\sigma_g$ and $V_{rot}$
at discrete redshifts sampling the range of observations in \citet{kas12}.  
We measure the intrinsic values of these quantities which are unaffected by observational effects such as seeing, slit width, 
and pixel scale.    The quantity $\sigma_g$ is measured as the average integrated velocity dispersion of the gas 
in the direction perpendicular to the disk, similar to observations.  
To measure $\sigma_g$, we step across the galaxies in a face-on position in 0.5 kpc radial bins and
measure the line-of-sight velocity dispersion in each.   The value of $\sigma_g$ is taken as
the mean value of the line-of-sight dispersions among the bins.
The rotation velocity $V_{rot}$ is taken as the maximum line-of-sight rotation velocity of the gas.
To measure $V_{rot}$, we step across the galaxies in an edge-on position in 0.5 kpc radial bins and
measure the line-of-sight rotation velocity in each.  We use the actual gas particle velocities 
rather than circular velocities since this provides the most direct comparison with observations.
From these velocity measurements, we create a rotation curve (rotation velocity versus radius).  
The maximum value of the rotation curve is adopted as $V_{rot}$,
similar to the observations which use the rotation velocity on the flat part of the rotation curve.

Our measurements are of intrinsic quantities and therefore are different from the mock observations 
of e.g., \citet{covi} which take into account myriad
observational effects.  Our goal is to determine the intrinsic kinematic evolution in the simulations,
not to investigate observational effects as in \citet{covi}.

Median values of $\sigma_g$ and $V_{rot}$ at discrete redshifts for the warm and cold gas in our 
4 simulated galaxies are shown in the top panels of Figure~\ref{settling}.  Values of these quantities
for the individual simulated galaxies are shown in the bottom panels.  In the top panels, the median values
are compared to those for an observed mass-limited sample of 270 star-forming galaxies from \citet{kas12}.
Qualitatively, the simulated galaxies follow
the same trends as the observations: they increase in $\sigma_g$ and decrease in $V_{rot}$ with
increasing redshift over $0.1<z<1.2$.  In other words, both
the simulations and the observations decrease in $\sigma_g$ and 
increase in $V_{rot}$ with time over the last $\sim8$ billion years,  {\it demonstrating that the simulated
galaxies undergo the disk settling found in observations.}  
Similarly, \citet{bird} find that the ratio of ordered to disordered motions ($V_{rot}/\sigma_g$) 
decreases with time for a similar mass galaxy from the Eris simulation.
Furthermore, the scatter in $\sigma_g$ and $V_{rot}$ for the warm/cold gas in the individual simulated galaxies
is large, similar to the scatter in the observations (Figure 5 in \citealt{kas12}).

As expected, the normalizations of the simulated and observed relations differ.    
The median values of $V_{rot}$ for the simulated galaxies are greater since they are more massive on average than 
the observed galaxies, and $V_{rot}$ generally scales with stellar mass for disk galaxies.  The masses of the simulated galaxies 
range $4.2-4.5\times 10^{10}$ M$_{\odot}$ versus $6.3 - 50.1 \times 10^9$ M$_{\odot}$ for the observations.
(The simulations and observations adopt \citealt{krou} and \citealt{chab} IMFs, respectively, which result in consistent
stellar masses.)  The normalization of the $\sigma_g$ versus $z$ relations for the warm and cold gas are higher and lower than the 
observations, respectively.  In the simulations the warm gas is affected by feedback from supernovae,
which results in higher $\sigma_g$ and leads to hotter temperatures compared to observed \ion{H}{2} regions.  
By definition, the cold gas is cooler than observed \ion{H}{2} regions, which results in lower $\sigma_g$.
In addition, we note that the median stellar mass of the mass-limited observational sample decreases with decreasing
redshift (Figure 4 in \citealt{kas12}).  Since the observed evolution is to increasing $V_{rot}$ and decreasing
$\sigma_g$ with decreasing redshift, this changing median mass makes these trends lower limits to the
intrinsic evolution.  The magnitude of this effect is unclear and we defer a more detailed analysis
of it to future comparisons with significantly larger samples of simulated galaxies.

\subsection{Quantitative Trends of $\sigma_g$ and $V_{rot}$ with Redshift}

Linear relations are fit to the median points in Figure~\ref{settling} by performing
least-squares fits which take into account the errors in $\sigma_g$ and $V_{rot}$
(errors in redshift are negligible).  When fitting, to avoid covariances,
we zero-point the medians near the middle of the samples such that they vary around 
$\sim$zero.  We obtain the following fits for $\sigma_g$
for the warm gas, cold gas, and observations, respectively:
\begin{alignat}{3}
\sigma_g - 50 &= (21.5 \pm 6.2)  &(z - 0.6) - (4.0 \pm 1.8),\\
\sigma_g - 20 &= (11.7 \pm 7.4)    &(z - 0.6) - (2.1 \pm 1.8),\\
\sigma_g  - 35 &= (26.0 \pm 5.7) &(z - 0.6) - (3.1 \pm 1.4).
\end{alignat}
These fits have $\chi^2$ values of 0.8, 1.3, and 4.7, respectively.
As for the observations, the median $\sigma_g$ of both the warm and cold gas grows with increasing redshift to $z=1.2$ 
(i.e., decreases with time), although the relation for the cold gas is consistent with no evolution.
The slopes of the relations for the warm gas and observations are consistent within uncertainties. 
As mentioned above, the normalization of the
relations for the warm and cold gas are higher and lower than the observations, respectively.

We obtain the following fits for $V_{rot}$ for the warm gas, cold gas, and observations, respectively:
\begin{alignat}{3}
V_{rot} - 175 &= (-56.9 \pm 22.0) (z - 0.6) - (7.9 \pm 6.2),\\
V_{rot} - 200 &= (-20.7 \pm 15.1) (z - 0.6) - (4.9 \pm 4.3),\\ 
V_{rot} - 100 &= (-48 \pm 11.5)     (z - 0.6) - (7.8 \pm 3.0).
\end{alignat}
These fits have $\chi^2$ values of 2.7, 1.5, and 9.2, respectively.
As for the observations, the median $V_{rot}$ of both the warm and cold gas decreases with
increasing redshift to $z=1.2$ (i.e., increases with time).
The simulated galaxies have a higher normalization (i.e., have faster median rotation velocities)
than the real galaxies, as expected because they are
on average more massive than the observed galaxy sample, as discussed above.

\subsection{Kinematics and Morphology}

As is the case for real galaxies over $0.1<z<1.2$ \citep[e.g.,][]{flor, kas7, yang, kas12}, the kinematics of 
simulated galaxies are reflected in their morphologies.
Images of two of our four simulated galaxies are shown in Figure~\ref{images} at discrete redshifts over the 
redshift range in Figure~\ref{settling}.  Both galaxies have more disturbed morphologies at $z=1.0$, and become
progressively less disturbed with decreasing redshift.  At $z=0$ they appear as ordered disk galaxies with
little disturbances or peculiarities. 
This morphological transformation is consistent with the kinematic settling they undergo
(i.e, decrease in $\sigma_g$ and increase in $V_{rot}$ with time).

The images in Figure~\ref{images} are produced by taking into account the effects of radiative transfer and dust
using a Monte-Carlo ray tracing program designed to pair with hydrodynamic
simulations ({\tt Sunrise};  \citep{sunrise1, sunrise2}.  In short, {\tt Sunrise} creates a spectral energy distribution 
for each star particle in the simulation based on its
age and metallicity, using the {\tt Starburst99} stellar population modeling software \citep{leit}.
The metallicities of the gas particles determine how light from the galaxy is attenuated by dust,
and a constant dust to gas ratio of 0.4 is assumed.  The resulting images are then convolved with the SDSS $g$, $r$, and 
$i$ band filters to produce those in Figure~\ref{images}.

\section{Conclusions}

For the first time we measure the {\it evolution} of the kinematics of a suite of simulated star-forming galaxies over the last
$\sim8$ billion years, over nearly half of the age of the universe.
Specifically, we measure the evolution of disordered motions ($\sigma_g$) and ordered rotation ($V_{rot}$) 
of gas in a suite of four simulated galaxies.
Measurements are compared with recent observations which show the progressive 
settling of gas in disk galaxies over $0.1<z<1.2$ from 
disordered systems into the ordered disk galaxies common today \citep{kas12}.  
We find that the simulated galaxies follow the same trends as the observations:
they progressively decrease in disordered motions ($\sigma_g$) and increase in ordered rotation ($V_{rot}$) with 
time.  The scatter in $\sigma_g$ and $V_{rot}$ among the 4 simulated galaxies is also similar to that 
found for the observations.
Differences in normalization between the observations and simulations 
can be attributed to differences in the average stellar mass of the galaxy samples and gas temperatures/regions
probed.  A larger sample of simulated galaxies by at least an order of magnitude is needed to
confirm our findings.   However, it is encouraging that with the simulations we have at hand,
we find similar trends and scatter as the observations.

Reproducing these observations is an important 
benchmark for any theory or simulation of disk galaxy formation.  The observations
trace the internal kinematics of star-forming galaxies over a significant period of time.
Internal kinematics directly dictate how the gas in galaxies is arranged, and are strongly affected by 
physical processes such as feedback from star-formation,
major/minor mergers, and smooth accretion of baryons.
Our next step in a future work is to place constraints on the processes which
have the most direct effect on disk settling.

\acknowledgments
FG acknowledges support from NSF grant AST-0607819.  Resources supporting this work were 
provided by the NASA High-End Computing (HEC) Program through the NASA Advanced Supercomputing (NAS)
Division at Ames Research Center.  AB would like to thank Jay Gallagher for helpful conversations.

\end{document}